# Multi-Optical Comb Metrology using Relative Carrier Envelope Phase Control and Demonstration for Arbitrary Polarization Modulation

Akifumi Asahara,[1,2] and Kaoru Minoshima[1,2,a]

[1]*Department of Engineering Science, Graduate School of Informatics and Engineering, The University of Electro-Communications (UEC), 1-5-1 Chofugaoka, Chofu, Tokyo 182-8585 Japan*

[2]*JST, ERATO MINOSHIMA Intelligent Optical Synthesizer (IOS) Project, 1-5-1 Chofugaoka, Chofu, Tokyo 182-8585 Japan*

a) Author to whom correspondence should be addressed. Electronic mail: k.minoshima@uec.ac.jp

## ABSTRACT

We propose a novel measurement technique that exploits the high coherent controllability of multi-comb systems, which corresponds to a generalization of the frequency control between the comb sources. In this paper, we particularly focus on the arbitrary relative carrier envelope phase (CEP) control through the relative offset frequency of two combs, $\Delta f_{\mathrm{ceo}}$. We successfully demonstrate polarization-modulated comb generation and its coherent detection in the developed system. The proof-of-principle experiment indicates the potential of multi-comb systems, enabling a rapid, precise, and arbitrary coherent modulation method utilized for a wide variety of metrological applications.

## MANUSCRIPT

1. Introduction

The optical frequency comb is a well-known fundamental technique in the field of metrology [1] because it exhibits excellent performance in terms of frequency accuracy, coherence, and controllability. By stabilizing the two frequencies of an optical comb source, $f_{\text{rep}}$ and $f_{\text{ceo}}$, the comb can be treated as a precise and useful "frequency ruler," where the absolute frequency of each longitudinal mode is accurately determined. Here, $f_{\text{rep}}$ and $f_{\text{ceo}}$ denote the mode separation (and the repetition rate) and the carrier envelope offset frequency of the comb, respectively. It is also well known that in the time domain, $f_{\text{ceo}}$ relates to the pulse-to-pulse variation of the carrier envelope phase (CEP), $\phi_{\text{CEP}}^{\text{p-p}}$, through the following relationship: $\phi_{\text{CEP}}^{\text{p-p}} = 2\pi f_{\text{ceo}}/f_{\text{rep}}$ [2]. This implies that optical frequency combs have high coherent controllability, which can be achieved only by electronically controlling the frequency parameters.

In multi-comb measurements, it is notable that not only $f_{\text{rep}}$ and $f_{\text{ceo}}$, but also $\Delta f_{\text{rep}}$ and $\Delta f_{\text{ceo}}$ are significant parameters, where $\Delta f_{\text{rep}}$ and $\Delta f_{\text{ceo}}$ denote the difference in $f_{\text{rep}}$ and $f_{\text{ceo}}$ between the comb sources of interest, respectively. $\Delta f_{\text{rep}}$ controls the relative temporal delay between the pulses emitted from the two combs. On the other hand, $\Delta f_{\text{ceo}}$ characterizes the relative phase between the pulse-train pairs of the two combs. When $\Delta f_{\text{rep}}$ is zero, the pulse-to-pulse variation of the relative CEP is expressed as $\Delta \phi_{\text{CEP}}^{\text{p-p}} = 2\pi \Delta f_{\text{ceo}}/f_{\text{rep}}$, which can be generally described as a function of time, $\Delta \phi_{\text{CEP}}(t) = 2\pi \Delta f_{\text{ceo}} t$. This indicates that the $\Delta f_{\text{ceo}}$ control in a multi-comb system enables a rapid, precise, and arbitrary modulation for the relative CEP of pulse-train pairs. In a multi-comb system, a novel and versatile light source with a high coherent controllability can be constructed by utilizing the relationships between the frequencies, $f_{\text{rep}}$, $f_{\text{ceo}}$, $\Delta f_{\text{rep}}$, and $\Delta f_{\text{ceo}}$.

Such controllability has been partially utilized in some advanced measurements. In particular, an asynchronous optical sampling (ASOPS) method, which is realized using two synchronized optical comb sources with a slight difference in the repetition frequencies, $\Delta f_{\text{rep}}$, is known as a powerful scheme that achieves rapid measurements over a wide dynamic range. This technique has been applied for pump–probe measurements [3,4] and terahertz spectroscopy [5], achieving a wide temporal dynamic range. Recently, the technique has also been applied to perform Fourier measurements in the dual-comb spectroscopy (DCS) [6,7], achieving a high frequency resolution of the order of several tens of megahertz or less with a very short acquisition time. At the early stage of the DCS study, it was developed mainly in the field of precision spectroscopy for molecular gas samples [7]. However, since then, its applicability has rapidly extended to other fields such as nonlinear spectroscopy [8–10], the characterization of solid materials [11,12], and ultrafast time-resolved spectroscopy [13].

In the conventional DCS, the high frequency controllability has been limitedly utilized in certain conditions. The "coherent averaging" technique [14], where $f_{\text{rep}}/\Delta f_{\text{rep}}$ and $\Delta f_{\text{ceo}}/\Delta f_{\text{rep}}$ are set as integers ($\Delta f_{\text{ceo}}$ is typically set to zero), are used to stabilize the timing and the phase of the observed interference waveform. In other cases, by setting the $f_{\text{rep}}/\Delta f_{\text{rep}}$ as a quasi-integer, a temporal data interleaving was performed [15]. However, to the best of our knowledge, there have been no further advanced frequency control schemes including arbitrary $\Delta f_{\text{ceo}}$ control. In the present study, we fully generalize the relationships between the frequencies, $f_{\text{rep}}$, $f_{\text{ceo}}$, $\Delta f_{\text{rep}}$, and $\Delta f_{\text{ceo}}$, and attempt to bring out the potential ability of multi-comb systems to the utmost limit.

In this study, we propose a novel measurement technique that exploits the high coherent controllability of multi-comb systems. The concept is a generalization of the frequency control in coherently linked optical combs, including the relative CEP control. Here, we consider the multi-comb system as a novel light source with high coherent controllability. To present the usefulness of this concept, we demonstrate a proof-of-principle experiment of polarization-modulated (PM) comb generation using the relative CEP control and its coherent detection, which were performed by three phase-linked comb lights. This study will pave the way for the development of future advanced coherent measurement methods using optical frequency combs.

2. Experimental

We demonstrated the PM comb generation and detection to present the usefulness of its high coherent controllability using the arbitrary $\Delta f_{\text{ceo}}$ control, which has not been widely exploited in the conventional DCS experiments. First, the concept of this experiment is as follows. When two orthogonally polarized comb pulses with the same $f_{\text{rep}}$ are synthesized as shown in Fig. 1(a), the polarization of the interfered field will be changed to linear, elliptic, and circular states depending on the temporal evolution of the relative CEP. As aforementioned, because the temporal variation of the relative CEP can be arbitrarily modulated according to the $\Delta f_{\text{ceo}}$ control, an arbitrary polarization control can be achieved. The polarization modulation itself has previously been performed using electro-optic devices such as photo-elastic modulators [16]. However, by newly introducing the sophisticated comb technology, it is expected to be improved in terms of the speed, accuracy, and arbitrariness of the polarization control.

The developed system is schematically shown in Fig. 1(c), and is generally composed of two parts, i.e., PM comb generation and coherent detection. In the generation part, we employed a lab-built Er fiber comb as the light source, where $f_{\text{rep}}$ was ~56.5 MHz and the center wavelength was ~1,560 nm (~192 THz). Both $f_{\text{rep}}$ and $f_{\text{ceo}}$ were phase-stabilized based on a 10-MHz microwave reference; see our previous reports for details of the locking scheme [11,13]. The spectrum was cut by a 1-nm bandpass filter (not shown) to prevent polarization dispersion. The comb light was incident to an acousto-optic modulator (AOM) exploited as a frequency shifter (Model 3165-1, Crystal Technology), and it was split into zeroth and first order diffracted comb light. Then, as shown in Fig. 1(b), $\Delta f_{\text{ceo}}$ was induced between these two combs as $\Delta f_{\text{ceo}} = f_{\text{AOM}} - n f_{\text{rep}}$, where $f_{\text{AOM}}$ is the induced frequency to the AOM and $n$ is an integer. The polarizations of the two comb lights were adjusted to be orthogonal using waveplates, and the pulses were spatially and temporally overlapped at a polarization beamsplitter. The initial relative phase was fine-tuned by a mechanical delay line located in the optical path of the zeroth beam. As the synthesized light, the PM comb, with a modulation rate of $\Delta f_{\text{ceo}}$, was coherently generated.

To utilize the PM comb, as naturally expected, it will be effective to adopt some modulation measurement method, which is synchronized to the polarization variation. Here, we adopted DCS as such a synchronized monitoring method, and attempted the coherent detection of the PM comb. To achieve this experiment, we also employed another Er fiber comb as the local comb, which had similar properties as the first Er comb. The repetition rate was controlled to be slightly different from the first Er comb as $f_{\text{rep}} + \Delta f_{\text{rep}}$, and the $f_{\text{ceo}}$ was set to be the same as that of the zeroth comb. The two Er fiber combs were phase-locked via a slave continuous-wave laser (not shown), and tight stabilization was achieved so that the relative linewidth was much narrower than 1 Hz; see our previous reports for details [11,13]. The PM comb was spatially overlapped with the local comb, which had a 45°-inclined linear polarization to the ground, using a beamsplitter. The interference signal, i.e., interferogram (IGM), was obtained using InGaAs detectors (bandwidth: 100 MHz) and a digitizer (resolution: 14 bits, bandwidth: 100 MHz) with a sampling rate of $f_{\text{rep}} + \Delta f_{\text{rep}}$. 30-MHz low pass filters were used to satisfy the Nyquist condition. By detecting both of vertical and horizontal IGMs simultaneously, the information of the polarization of the PM comb was directly obtained as three-dimensional (3D) IGMs.

Furthermore, to coherently detect the IGMs, we synchronized the two frequencies, $\Delta f_{\text{ceo}}$ (difference in $f_{\text{ceo}}$ between the zeroth and first diffracted combs generated by the AOM) and $\Delta f_{\text{rep}}$ (difference in $f_{\text{rep}}$ between the two Er combs). As mentioned above, the polarization of the PM comb was modulated with a rate of $\Delta f_{\text{ceo}}$. On the other hand, in DCS, the IGM is obtained with a sampling rate of $\Delta f_{\text{rep}}$. Therefore, by synchronizing the them, the polarization corresponding to each IGM can be precisely controlled. By varying the relationship, a wide variety of coherent detections can be achieved, which can also be interpreted as the generalization of the "coherent averaging" scheme. In this study, we performed the experiment by changing the ratio between $\Delta f_{\text{ceo}}$ and $\Delta f_{\text{rep}}$.

3. Results and Discussion

First, we show the results obtained with the condition $\Delta f_{\text{ceo}} = \Delta f_{\text{rep}}/4$ ($f_{\text{rep}} = $ ~56.5 MHz, $\Delta f_{\text{rep}} = $ ~1002 Hz, $f_{\text{ceo}} = $ ~21.34 MHz, and $\Delta f_{\text{ceo}} = $ ~250 Hz ($f_{\text{AOM}} = \Delta f_{\text{ceo}} + 3f_{\text{rep}} = $ ~169.5 MHz)), where these frequencies were precisely controlled using the accuracy of the microwave reference. In this case, the relative CEP, $\Delta \phi_{\text{CEP}}(t)$, between the pulses constructing the PM comb changes by $\pi/2$ for every IGM signal as shown in Fig. 2(a). As a result, the four-fold IGMs were observed in the vertical and horizontal axes as shown in Fig. 2(b), where we can find the characteristic changes in the amplitudes of the center burst signals depending on the polarization states of the PM comb. Furthermore, as shown in Fig. 2(c), the 3D IGMs were reconstructed from these data. Here, we could clearly observe the successful periodic polarization modulation with horizontal, counterclockwise, vertical, and clockwise polarization states. In this way, we confirmed that the characteristic PM comb generation was achieved by the relative CEP control and that its coherent detection was also achieved using the extended DCS scheme.

Here, we observed the waveforms of the four-fold IGMs were coherently integrated because the frequency parameters were set to inherently satisfy the coherent averaging condition. With an additional real-time computational phase compensation method, which was also adopted in our previous studies [11,13], the persistent coherent data integration could be achieved. The IGMs in the present case were obtained with a 3-s integration, and the signal-to-noise ratio will be further improved with a longer acquisition.

We also found that the visibility of the polarization interference appears not to be perfect. We speculate that this was mainly attributed to the discrepancy of the spatial beam pattern between the zeroth and first diffracted light from the AOM. Such discrepancies can be improved using two spectrally designed and phase-linked comb sources without AOMs.

With other frequency setups, the behavior of the IGMs of the PM comb drastically changes. For example, with a condition of $\Delta f_{\text{ceo}} = \Delta f_{\text{rep}}/2$ (= ~501 Hz), "polarization switching" can be realized, because the relative CEP of the PM comb causes a π-shift for each IGM. Figures 3(a) and 3(b) show the magnified 3D IGMs when the initial relative CEP was set to be (a) zero and (b) π/2, which was controlled by the delay line. Here, it is found that the linear and circular polarization switching for each IGM could be controlled by changing the initial relative CEP.

Moreover, when $\Delta f_{\text{ceo}}$ is synchronized to $f_{\text{rep}}$, a much faster modulation can be achieved. For example, with a condition of $\Delta f_{\text{ceo}} = f_{\text{rep}}/2$ (= ~28.3 MHz), "pulse-to-pulse polarization switching" can be realized. In this case, the data points corresponding to each polarization are sampled alternatingly, and two 1/2-downsampled IGMs are simultaneously obtained as a temporally interleaved waveform. Fig. 4(a) shows the IGMs obtained when the initial relative CEP was set to be about π/2. By alternatingly resampling the points, the 1/2-downsampled 3D IGMs could be obtained, as shown in Figs. 4(b) and 4(c). From these results, we can clearly find that the pulse-to-pulse polarization switching in the inverse circular directions was realized in the PM comb.

We expect that the characteristic PM comb generation and its coherent detection techniques will be useful for a wide variety of potential applications. For example, rapid polarization modulation and switching can be used in spectroscopic studies. Together with the concept of the well-developed polarization modulation spectroscopy [16], optical properties of materials, such as birefringence, circular dichroism, optical chirality, or the properties of topological insulators, can be effectively characterized. Furthermore, the high coherent controllability of the relative CEP can also be utilized as an advanced light source for optical manipulation using optical lattices [17,18]. Because the spatial interference pattern of the optical lattice was determined by the relative CEP of the overlapped combs, the controllability enables a rapid, precise, and arbitrary spatial intensity modulation. We believe that concepts such as these will enhance the potential of multi-comb systems as advanced coherent light sources that are utilized for versatile applications.

4. Conclusion

In this study, we proposed a novel measurement technique using the excellent coherent controllability of multi-comb system, which corresponds to a generalization of the frequency control between the comb sources. To demonstrate the usefulness of this concept, the PM comb generation and its coherent detection were successfully achieved by the developed system through a precise control of the frequency parameters of the combs, $f_{\text{rep}}$, $f_{\text{ceo}}$, $\Delta f_{\text{rep}}$, and $\Delta f_{\text{ceo}}$. Although the several frequency configurations were selected in this case, the condition can be fully arbitrarily designed depending on the experiments of interest. This proof-of-principle study paves the way for novel applications of multi-comb systems, enabling a rapid, precise, and arbitrary coherent modulation that can be utilized in a wide variety of metrological studies.

**ACKNOWLEDGMENTS**

This work was supported by the JST ERATO MINOSHIMA Intelligent Optical Synthesizer Project (Grant Number JPMJER1304) and JSPS Grant-in-Aid for Young Scientists (B) (Grant Number JP17K14322).

**FIGURES**

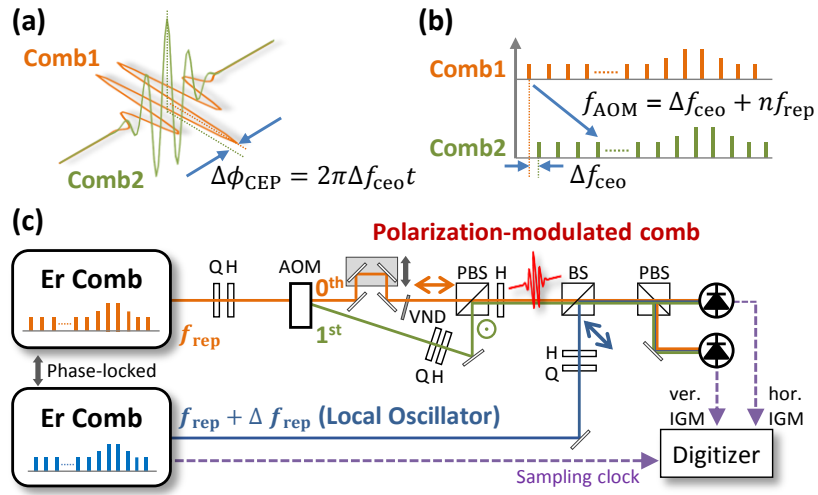

Fig. 1. (a) Schematic of orthogonally polarized comb pulses used to generate the PM comb. (b) Spectra of the two combs with $\Delta f_{\text{ceo}}$. (c) Developed system for the PM comb generation and detection experiment. AOM: acousto-optic modulator; Q: quarter waveplate; H: half waveplate; BS: beamsplitter; PBS: polarization beamsplitter; IGM: interferogram.

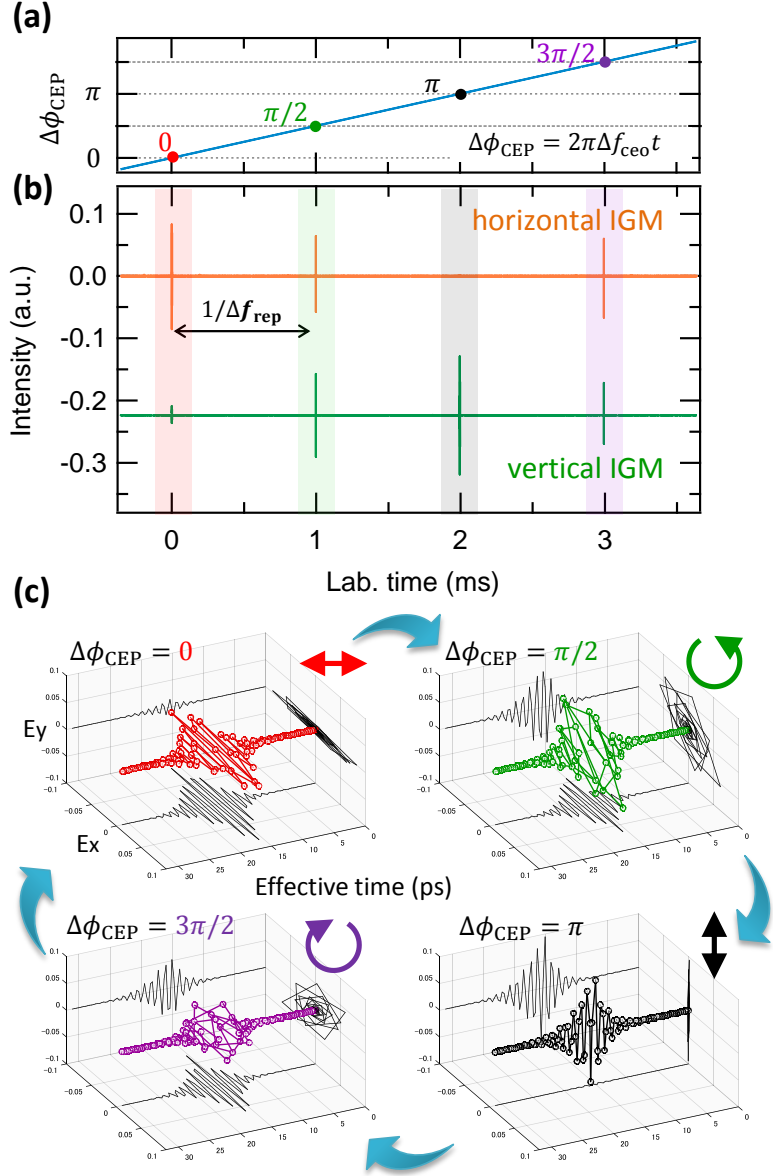

Fig. 2. Results obtained when $\Delta f_{\text{ceo}} = \Delta f_{\text{rep}}/4$ (= ~250 Hz). (a) Schematic of the temporal evolution of the relative CEP, $\Delta\phi_{\text{CEP}}(t)$. (b) Simultaneously obtained IGMs along the horizontal and vertical axes. (c) 3D IGMs constructed from the data in (b), which were magnified around the center burst signals (around masked area in (b)). Colored arrows denote the polarization states of the synthesized pulses.

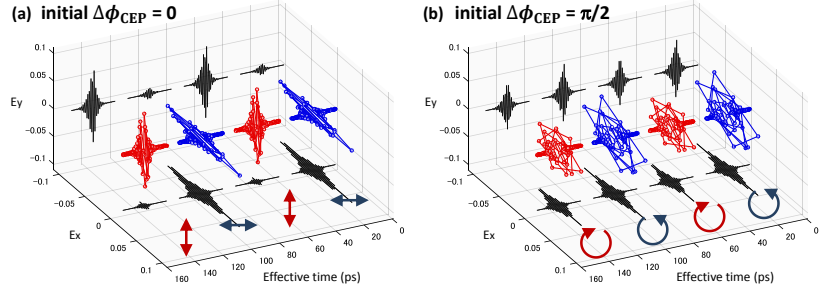

Fig. 3. 3D IGMs magnified around the center bursts, which were observed when $\Delta f_{\text{ceo}} = \Delta f_{\text{rep}}/2$ (= ~501 Hz). (a) Linear and (b) circular polarization switching were controlled by changing the initial relative CEP.

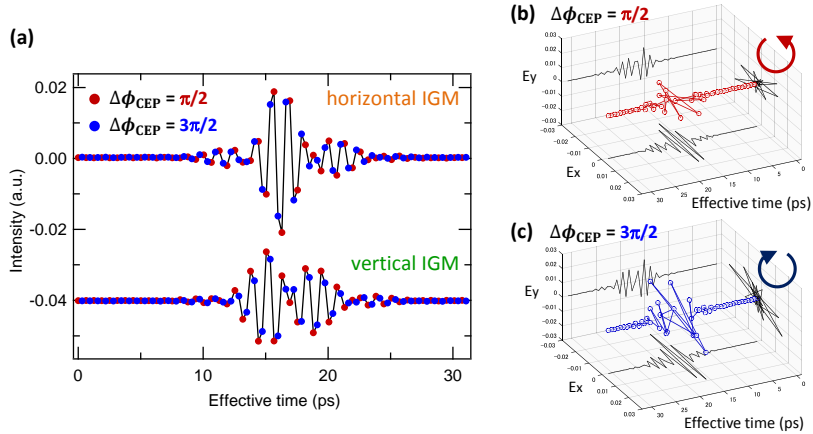

Fig. 4. Results observed when $\Delta f_{\text{ceo}} = f_{\text{rep}}/2$ (= ~28.3 MHz). (a) IGMs along the horizontal and vertical axes, comprising two 1/2-downsampled IGMs (red and blue) simultaneously obtained as the temporally interleaved waveforms. Resampled 1/2-downsampled 3D IGMs with $\Delta \phi_{\text{CEP}}$ of (b) π/2 and (c) 3π/2, indicating the achievement of pulse-to-pulse circular polarization switching in the PM comb.

13